\documentclass[useAMS,usenatbib]{mn2e}
\usepackage{txfonts}
\usepackage{graphics}
\usepackage{times}

\def\del#1{{}}

\newcommand{\apj}{ApJ}

\newcommand{\mnras}{MNRAS}

\newcommand{\nat}{Nature}

\def\be{\begin{equation}}
\def\ee{\end{equation}}
\def\ba{\begin{eqnarray}}
\def\ea{\end{eqnarray}}

\newcommand{\hompc}{\,h\,{\rm Mpc}^{-1}}

\newcommand{\simgt}%
{\,\hbox{\lower0.6ex\hbox{$\sim$}\llap{\raise0.6ex\hbox{$>$}}}\,}
\newcommand{\simlt}%
{\,\hbox{\lower0.6ex\hbox{$\sim$}\llap{\raise0.6ex\hbox{$<$}}}\,}

\begin{document}

\title[Redshift-space distortions]{Forecasting Cosmological Constraints from
Redshift Surveys}

\author[White et al.]{
\parbox{\textwidth}{
Martin White$^{1}$, Yong-Seon Song$^{2}$ and Will J. Percival$^{2}$}
\vspace*{4pt} \\
$^{1}$ Departments of Physics and Astronomy, University of California,
Berkeley, CA 94720, USA\\
$^{2}$ Institute of Cosmology and Gravitation, University of
Portsmouth, Portsmouth, P01 2EG, UK}

\date{\today} 
\pagerange{\pageref{firstpage}--\pageref{lastpage}}

\maketitle
\label{firstpage}

\begin{abstract}
  Observations of redshift-space distortions in spectroscopic galaxy
  surveys offer an attractive method for observing the build-up of
  cosmological structure, which depends both on the expansion rate of
  the Universe and our theory of gravity. In this paper we present a
  formalism for forecasting the constraints on the growth of structure
  which would arise in an idealized survey.  This Fisher matrix based 
  formalism can be used
  to study the power and aid in the design of future surveys.
\end{abstract}


\section{Introduction}

The growth of large-scale structure, as revealed in the clustering of
galaxies observed in large redshift surveys, has historically been one
of our most important cosmological probes.  This growth is driven by a
competition between gravitational attraction and the expansion of space-time,
allowing us to test our model of gravity and the expansion history of the
Universe.  Despite the fact that galaxy light doesn't faithfully trace the
mass, even on large scales, galaxies are expected to act nearly as test
particles within the cosmological matter flow.  Thus the motions of galaxies
carry an imprint of the rate of growth of large-scale structure and allows
us to probe both dark energy and test General Relativity
\cite[e.g.][for recent studies]{jain08,Song08a,Song08b,PerWhi08,McDSel08}.

This measurement of the growth of structure relies on redshift-space
distortions seen in galaxy surveys \citep{Kai87}.  Even though we expect the
clustering of galaxies in real space to have no preferred direction, galaxy
maps produced by estimating distances from redshifts obtained in spectroscopic
surveys reveal an anisotropic galaxy distribution.
The anisotropies arise because galaxy recession velocities, from which
distances are inferred, include components from both the Hubble flow and
peculiar velocities driven by the clustering of matter
\citep[see][for a review]{HamiltonReview}.
Measurements of the anisotropies allow constraints to be placed on the
rate of growth of clustering.

Ever larger surveys have provided ever tighter constraints.
Analyses using the 2-degree Field Galaxy Redshift Survey
\citep[2dFGRS;][]{colless03} have measured redshift-space distortions
in both the correlation function \citep{peacock01,hawkins03} and power
spectrum \citep{percival04}.
Using the Sloan Digital Sky Survey \citep[SDSS;][]{york00}, redshift-space
distortions have also been measured in the correlation function
\citep{Zeh05,Oku08,cabre08}, and using an Eigenmode decomposition to separate
real and redshift-space effects \citep{tegmark04,tegmark06}.
These studies were recently extended to $z\simeq1$ \citep{guzzo08} using the
VIMOS-VLT Deep Survey \citep[VVDS;][]{lefevre05,garilli08}.
In addition to measuring clustering growth at $z=0.8$, this work has emphasized
the importance of using large-scale peculiar velocities for constraining models
of cosmic acceleration.
Current constraints on the growth rate are at the several tens of percent
level \citep[e.g.][]{NesPer08,Song08b}, but observational progress is rapid.

In the next section we shall outline the formalism for forecasting constraints
on cosmological quantities from measurements of redshift space distortions,
and compare it with previous forecasts.
We begin with the simplest model and then investigate various refinements.
We finish in \S\ref{sec:conclusions} with a discussion of future directions.
For illustration we shall assume a fiducial $\Lambda$CDM cosmology with
$\Omega_{\rm m}=0.25$, $h=0.72$, $n=0.97$ and $\sigma_8=0.8$ (in good
agreement with a variety of observations) when computing specific predictions
for future surveys.

\section{The Fisher matrix}  \label{sec:Fisher}

The Fisher matrix provides a method for determining the sensitivity of a
particular measurement to a set of parameters and has been extensively used
in cosmological forecasting and optimization.  Here we adapt this methodology
to our particular problem.

\subsection{The simplest case}  \label{sec:fish_simple}

Under the assumption that the density field has Gaussian statistics and
uncorrelated Fourier modes, the Fisher matrix for a set of parameters
$\{p_i\}$ is \citep[e.g.][]{Teg98}
\begin{equation}
  F_{ij} = \frac{1}{2}\int \frac{d^3k}{(2\pi)^3}
  \left( \frac{\partial \ln P}{\partial p_i} \right)
  \left( \frac{\partial \ln P}{\partial p_j} \right)
  V_{\rm eff}\left(\vec{k}\right)
\label{eqn:fisher}
\end{equation}
where $P$ is the power spectrum and the mode counting is determined by
the effective volume \citep{FKP}
\begin{equation}
  V_{\rm eff}\left(\vec{k}\right) \equiv
  V_0\left( \frac{\bar{n}P}{1+\bar{n}P} \right)^2
\end{equation}
which depends on the geometric volume of the survey, $V_0$, and the
number density, $\bar{n}$, of the tracer.  If $\bar{n}$ is high enough
then $V_{\rm eff}\simeq V_0$.  The constraints are dominated by regions
where $\bar{n}P\ge 1$, so it is safe to neglect the higher order
(in $\bar{n}^{-1}$) terms which arise assuming that galaxies are a
Poisson sample of the underlying density fluctuations
\citep{MeiWhi99}.

The simplest model for the observed galaxy distribution
is a linear, deterministic, and scale-independent galaxy bias,
with redshift space distortions due to super-cluster infall \citep{Kai87}
and no observational non-idealities.  In this case
$P_{\rm obs}\propto  \left(b+f\mu^2\right)^2 P_{\rm lin}(k)$
where $P_{\rm lin}$ is the linear theory mass power spectrum in real space,
$b$ is the bias and $\mu$ the angle to the line-of-sight.  The quantity of
most interest here is $f\equiv d\ln D/d\ln a$, the logarithmic derivative of
the linear growth rate, $D(z)$, with respect to the scale factor
$a=(1+z)^{-1}$.  In general relativity $f\approx \Omega_{\rm mat}(z)^{0.6}$
\citep[e.g.][]{Pee80}, while in modified gravity models it can be smaller by
tens of percent \citep[e.g.][figure 1]{Song08b}.
Redshift space distortions allow us to constrain $f$ times the 
normalization of the power spectrum (e.g.~$f(z)\sigma_8(z)$), or $dD/d\ln a$.
The derivatives in Eq.~(\ref{eqn:fisher}) are particularly simple
\begin{equation}
  \frac{\partial\ln P}{\partial b} = \frac{2}{b+f\mu^2}
  \qquad {\rm and} \qquad
  \frac{\partial\ln P}{\partial f} = \frac{2\mu^2}{b+f\mu^2}
  \qquad ,
\end{equation}
independent of the shape of the linear theory power spectrum, and hence of the
spectral index and transfer function.
Since we hold the normalization of the power spectrum fixed for these
derivatives, the fractional error on $f(z)\sigma_8(z)$ is equal to that on
$f$ in our formalism.
The errors on $b$ and $f$ depend sensitively on the maximum $k$ in the integral
of Eq.~(\ref{eqn:fisher}).  Since we are using linear theory we choose to cut
the integral off at $k\simeq 0.1\,h\,{\rm Mpc}^{-1}$ for our fiducial
cosmology.  This is close to the value at which \citet{PerWhi08} saw
departures from linear theory.

The bias and $f$ turn out to be anti-correlated, with a correlation
coefficient of $70-75\%$, depending on the precise sample.  We marginalize
over $b$ by first inverting the Fisher matrix to find the covariance matrix
and hence the error $f$, specifically
$\delta f=\left(F^{-1}\right)_{ff}^{1/2}$.  Hereafter we shall drop the
parentheses and write $F^{-1}_{ff}$ for the $ff$ component of $F^{-1}$.

Forecasts on $\delta f/f$ in this model can be regarded as an ``optimistic''
estimate of the reach of future observations, and the results are quite
encouraging.  For a $10\,(h^{-1}{\rm Gpc})^3$ survey\footnote{Out to $z=1$
over $10,000$ square degrees the (comoving) volume in our fiducial cosmology
is $14\,(h^{-1}{\rm Gpc})^3$.} at $z=0$ an unbiased population of tracers
with $\bar{n}P\gg 1$ would achieve $\delta f/f=1.6\%$, with the error scaling
as $V_0^{-1/2}$.
The constraint weakens with increasing bias, being $3\%$ for $b=2$ and $6\%$
for $b=4$.  The constraint also weakens as we reduce $\bar{n}$.
For $b=1$ and $\bar{n}=10^{-3}\,h^3\,{\rm Mpc}^{-3}$ the constraint is $1.8\%$,
increasing to $1.9\%$ for $\bar{n}=4\times 10^{-4}\,h^3\,{\rm Mpc}^{-3}$,
$2\%$ for $\bar{n}=2\times 10^{-4}\,h^3\,{\rm Mpc}^{-3}$ and
$3\%$ for $\bar{n}=10^{-4}\,h^3\,{\rm Mpc}^{-3}$.  The shot noise,
$\bar{n}P(0.1\,h\,{\rm Mpc}^{-1})\simeq 1$ for
$\bar{n}=2\times 10^{-4}\,h^3\,{\rm Mpc}^{-3}$, which explains the rapid
increase in $\delta f$ for $\bar{n}$ larger than this.

Conversely, increasing $k_{\rm max}$ to $0.2\,h\,{\rm Mpc}^{-1}$ reduces the
error on the $b=1$, $\bar{n}P\gg 1$ case to $0.6\%$.
Moving to higher redshift, keeping the bias fixed, makes the constraint
stronger as $f/b$ is increased.  By $z=1$, $f$ has increased to $0.83$ from
$0.44$ and $\delta f/f\simeq 1\%$ for our fiducial $b=1$, $\bar{n}P\gg 1$
example.  Note, however, that at higher $z$ the effects of shot noise would
typically be larger.

These constraints can be compared to the forecasts in \citet{guzzo08} who
present a fitting function for the relative error on $\beta=f/b$ of
\begin{equation}
  \frac{\delta\beta}{\beta} = \frac{50}{V^{1/2}\left(\bar{n}\right)^{0.44}}
\end{equation}
where $\bar{n}$ is measured in $h^3\,{\rm Mpc}^{-3}$ and $V$ in
$h^{-3}\,{\rm Mpc}^3$.  Note that both forecasts agree on the scaling with
volume, but the above scales approximately as the inverse square root
of the total number of galaxies in the survey and is independent of the
bias.
Taking into account the correlation between the constraints on $b$ and $f$
our forecast constraint is
\begin{equation}
  \frac{\delta\beta}{\beta} =
  b^{-1}\left[ \beta^2 F_{bb}^{-1}-2\beta F_{bf}^{-1}+F_{ff}^{-1}\right]^{1/2}
\end{equation}
Comparing our forecasts to this scaling we find relatively good agreement
for $b\simeq 1$ and $\bar{n}\simeq 10^{-4}\,h^3\,{\rm Mpc}^{-3}$, but the
\citet{guzzo08} scaling predicts much better constraints for higher number
density or more biased samples.

\subsection{Beyond linear theory} \label{sec:beyond}

Of course we do not expect the simple linear theory result with super-cluster
infall to be a perfect description of redshift space distortions on all scales.
Comparison with N-body simulations suggests that halos do closely follow the
matter velocity field and the major modification to the simple model at low
$k$ is in the quadrupole, with an additional effect coming from the generation
of multipoles higher than $4$.
By introducing more freedom into the model we will increase our ability to
describe the extra physics acting, and simultaneously begin to degrade our
sensitivity to $f$.

In \citet{PerWhi08} it was shown that a streaming model with a Gaussian
small-scale velocity provided an adequate fit to N-body simulations to
$k\simeq 0.1\,h\,{\rm Mpc}^{-1}$.  Under these assumptions one model for
the redshift space, galaxy power spectrum could be
\begin{equation}
  P_{\rm obs}(k,\mu) = \left(b+f\mu^2\right)^2 P_0(k) e^{-k^2\sigma_z^2\mu^2}
\end{equation}
where $P_0$ represents the mass power spectrum in real space and $\sigma_z$ is
to be regarded as a fit parameter which encompasses a variety of violations
of the traditional analysis.  We can additionally model inaccuracies in the
observed redshifts by a line-of-sight smearing of the structure.
In the limit that this smearing is Gaussian it can be absorbed into $\sigma_z$.

In this situation, the logarithmic derivatives with respect to $b$ and $f$ are
unchanged and the new derivative required is simply
$\partial\ln P/\partial\sigma_z^2=-k^2\mu^2$.
We now marginalize over $\sigma_z$ in addition to $b$ before reporting the
constraints on $f$.

Many of the trends with $b$ and $\bar{n}$ in our simple model also hold for
this extended model.
For our fiducial $10\,(h^{-1}{\rm Gpc})^3$ volume the constraint from our
$z=0$, unbiased tracers with $\bar{n}P\gg 1$ weakens from $\delta f/f=1.6\%$
to $3.2\%$ when marginalizing over $\sigma_z$.  The fiducial value of
$\sigma_z$ has little impact for $\bar{n}P\gg 1$.  However when
$\bar{n}P\simeq 1$ the error is increased from $3\%$ to $4\%$ to $20\%$
as $\sigma_z$ is increased from $0$ to $10\,h^{-1}$Mpc to $100\,h^{-1}$Mpc
for a sample with $b=1$.

Another alternative is to model the small-scale suppression with a Lorentzian,
which provides a better fit at higher $k$ and is a good match to the
superposition of Gaussians of different widths from halos of different masses
\citep{HaloRed}.  The two agree to lowest order in $k\sigma_z$, and which is
the regime of most interest here, with the Gaussian matching the results
of N-body simulations at small $k$ slightly better than the exponential
\citep{PerWhi08}.  Changing the form from Gaussian to exponential makes a
negligible change in our forecasts.

\subsection{Mode by mode} \label{sec:mode}

The forecasts above all made quite strong assumptions about the relationship
between the velocity and density power spectra, assumptions which are only
known to be true for quasi-linear scales within the context of General
Relativity.  The parameters $\{p_i\}$ in our Fisher matrix
(Eq.~\ref{eqn:fisher})
don't have to be cosmological parameters however.  We can fit directly for
the three independent power spectra (the density-density, velocity-velocity and
density-velocity spectra) rather than assuming that they are related by a
specific functional form \citep[e.g.][]{tegmark04}.
Such constraints would be applicable to a wide range of theories including
e.g., interacting dark energy, clustered dark energy or $f(R)$ gravity. 

\begin{figure}
\begin{center}
\resizebox{3.2in}{!}{\includegraphics{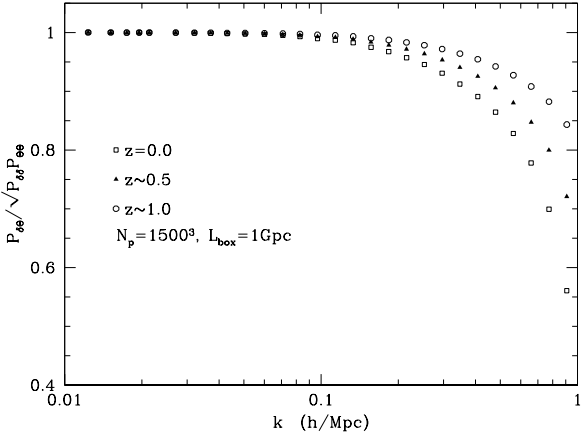}}
\end{center}
\caption{The correlation coefficient, $r(k)$,  between the density and velocity
divergence of the dark matter in an N-body simulation of a $\Lambda$CDM
cosmology, with the same cosmological parameters as our fiducial model.}
\label{fig:rk}
\end{figure}

\subsubsection{Correlation of $\delta$ and $\Theta$}

\begin{figure}
\begin{center}
\resizebox{3.2in}{!}{\includegraphics{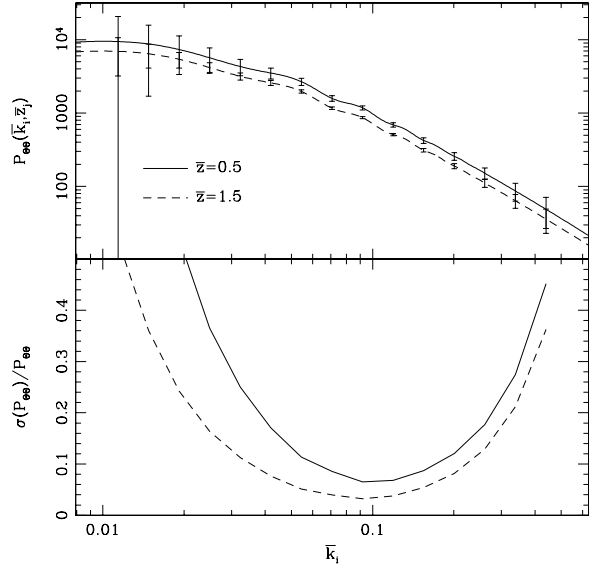}}
\end{center}
\caption{(Top) $P_{\Theta\Theta}(k_i,z_j)$ in redshift bins of width $\Delta z=0.2$
at $z_j=0.5$ (solid) and 1.5 (dashed) from a half-sky survey with $b=1.5$ and
$\bar{n} = 5\times 10^{-3}\,h^3\,{\rm Mpc}^{-3}$.
(Bottom) The fractional error on $P_{\Theta\Theta}(k_i,z_j)$ in the same bins
weighting modes with $\sigma_{\rm th}=0.1$ (see text).}
\label{fig:pktt}
\end{figure}

N-body simulations of $\Lambda$CDM cosmologies show that the density and
velocity divergence are highly correlated for $k<0.1\,h\,{\rm Mpc}^{-1}$
(see Fig.~\ref{fig:rk}) so we will begin by making the assumption that the
density and velocities are perfectly correlated
(to be relaxed in \S\ref{sec:decorrelation}).
Then the density-velocity cross-spectrum becomes the geometric mean of the
two auto-spectra and we have only two free functions.
If we write $\Theta$ for the velocity divergence in units of $aH$ the power
spectrum becomes
\begin{eqnarray}
P_{\rm obs}(k,\mu,z) &=& P_{gg}(k,z)
   + 2\mu^2\left[P_{gg}(k,z)P_{\Theta\Theta}(k,z)\right]^{1/2}\nonumber\\
  &+& \mu^4P_{\Theta\Theta}(k,z)
\end{eqnarray}
where $P_{gg}$ denotes the usual galaxy density auto-spectrum and we have
assumed that small-scale \citep[``finger of god'';][]{Jac72} effects have
been cleanly removed by e.g.~finger of god compression.
The parameters in the Fisher matrix, Eq.~(\ref{eqn:fisher}), are now the values
of the two spectra themselves, in bins of $k$ and $z$:
\begin{eqnarray}
  \frac{\partial \ln P_{\rm obs}(k_i,\mu,z_j)}{\partial P_{gg}(k_i,z_j)}
  &=& \frac{1}{P_{\rm obs}(k_i,\mu,z_j)}
  \left[1 + \mu^2
  \sqrt{\frac{P_{\Theta\Theta}(k_i,z_j)}{P_{gg}(k_i,z_j)}} 
  \right] \nonumber \\
  \frac{\partial\ln P_{\rm obs}(k_i,z_j)}{\partial P_{\Theta\Theta}(k_i,z_j)}
  &=&\frac{\mu^2}{P_{\rm obs}(k_i,\mu,z_j)}
  \left[\sqrt{\frac{P_{gg}(k_i,z_j)}{P_{\Theta\Theta}(k_i,z_j)}}+\mu^2\right]
\end{eqnarray}
and the variance of $P_{\Theta\Theta}(k_i,z_j)$ is given by
\begin{equation}
\sigma[P_{\Theta\Theta}(k_i,z_j)] = F_{22}^{-1}(k_i,z_j)
\end{equation}
where $F_{22}^{-1}(k_i,z_j)$ is 22-component of the inverse matrix of
$F_{\alpha\beta}$.
The constraint on $P_{\Theta\Theta}(k_i,z_j)$ in 2 redshift bins each of
$\Delta z=0.2$ is plotted in Fig.~\ref{fig:pktt} for a half-sky survey
with $\bar{n} = 5\times 10^{-3}\,h^3\,{\rm Mpc}^{-3}$ and $b=1.5$.

\subsubsection{Decorrelation of $\delta$ and $\Theta$} \label{sec:decorrelation}

The assumption of tight correlation between $\delta$ and $\Theta$ is a
reasonable one (densities grow where flows converge and velocities are high
where mass concentrations cause a large gravitational potential) but is
not required.  We can extend the formalism above by allowing the
cross-correlation coefficient,
\begin{equation}
  r(k) \equiv \frac{P_{g\Theta}}{
  \sqrt{P_{\Theta\Theta}(k)P_{gg}(k)} }
  \qquad ,
\end{equation}
to differ from unity.
The power spectrum can now be written in terms of 3 free functions
($P_{gg}$, $P_{\Theta\Theta}$, $r$) as
\begin{equation}
  P_{\rm obs} =
  \left(P_{gg}+2\mu^2\,r(k)\sqrt{P_{gg}P_{\Theta\Theta}}+
  \mu^4\,P_{\Theta\Theta}\right)\,G_{\rm FoG}(k,\mu;\sigma_z)  
    \label{eq:Pobs_tt}
\end{equation}
where $G_{\rm FoG}$ is a Gaussian describing the decrease in power
due to virial motions and the derivatives are given by
\ba
  \frac{\partial \ln P_{\rm obs}}{\partial P_{gg}}
  &=& \frac{1}{P_{\rm obs}}
  \left(1 + \mu^2r \sqrt{\frac{P_{\Theta\Theta}}{P_{gg}}}\right)
  \, G_{\rm FoG}\nonumber \\
  \frac{\partial \ln P_{\rm obs}}{\partial P_{\Theta\Theta}}
  &=& \frac{1}{P_{\rm obs}} \left(\mu^2r \sqrt{\frac{P_{gg}}{P_{\Theta\Theta}}}
  +\mu^4 \right)\, G_{\rm Fog}\nonumber \\
  \frac{\partial \ln P_{\rm obs}}{\partial r}
  &=& \frac{2}{P_{\rm obs}}\mu^2\sqrt{P_{gg}P_{\Theta\Theta}}\, G_{\rm FoG}
  \quad .
\label{eqn:mode_deriv}
\ea

We find that allowing $r(k)$ to be completely free degrades the constraint
on $P_{\Theta\Theta}$, and eventually $f$, significantly, until it is
equivalent to simply measuring the $\mu^4$ component in Eq.~(\ref{eq:Pobs_tt}).
To strengthen the constraint requires prior information about $r(k)$, which
can in principle be obtained from simulations or perturbation theory
calculations of structure formation in modified gravity models.
As an illustrative example, if we assume a prior
\begin{equation}
  \sigma_{\rm prior}(r)=1-r(k)
\end{equation}
where the error on $r$ is equal to its deviation from unity (using the
fiducial model with $r$ measured from N-body simulations as in
Fig.~\ref{fig:rk}) we find the constraint on $P_{\Theta\Theta}$ is almost
the same as we obtained before.

Also note that in this analysis we can mitigate our uncertainty in the
form of the small-scale redshift space distortion by downweighting modes
for which the fingers-of-god \citep{Jac72} are expected to be large.
The residual uncertainty after the weighting is
\begin{equation}
  F_{ij} = \frac{1}{2}\int \frac{V_0\,d^3k}{(2\pi)^3}
  \left( \frac{\partial \ln P}{\partial p_i} \right)
  \left( \frac{\partial \ln P}{\partial p_j} \right)
  V_{\rm eff}\left(k,\mu\right)w_{\rm FoG}(k,\mu)
\label{eqn:fisher_w}
\end{equation}
where the weight function $w_{\rm FoG}(k,\mu)$ could, for example, be given by
\begin{equation}
  w_{\rm FoG}(k,\mu)=
  \exp{\left[-\frac{(G_{\rm FoG}-1)^2}{\sigma_{\rm th}^2}\right]}
\end{equation}
where $G_{\rm FoG}$ is the finger-of-god suppression factor and
$\sigma_{\rm th}$ is a threshold value indicating our confidence in
the accuracy of the FoG model.

\subsection{Multiple populations}  \label{sec:fish_multiple}

Until now we have implicitly assumed that we are dealing with a single
population of objects.  However galaxies come in a variety of sizes,
luminosities, masses and types which exhibit different clustering patterns
but all of which are expected to respond to the same large-scale velocity
field.  \citet{McDSel08} pointed out recently that this allows, in principle,
for significant gains in determination of the growth of structure.  In fact,
in the limit of Gaussian statistics, perfectly deterministic bias and
infinitely dense tracers, one can measure the velocity power spectrum limited
only by the total number of modes in the survey in all directions.

To include multiple populations in the Fisher matrix approach, there
are two obvious ways of proceeding.  \citet{McDSel08} assumed that the
densities, $\delta_i$ for $1\le i\le N$, are the measured quantities and
built a covariance matrix in terms of the power spectra,
$\langle\widehat{\delta}_i\widehat{\delta}_j\rangle$, where a superscript
$\widehat{\phantom{\delta}}$ denotes a measured quantity that includes a
noise term.
An alternative and complementary approach is to extend the analysis presented
in \S\ref{sec:fish_simple} assuming that the power spectra are the measured
quantities.  For Gaussian fluctuations, in which all of the cosmological
information is encoded in the power spectrum, these approaches turn out to
be equivalent\footnote{We verified this by explicit numerical computation of
the Fisher matrices.}.  We develop this second approach here.

\begin{figure}
\begin{center}
\resizebox{3.1in}{!}{\includegraphics{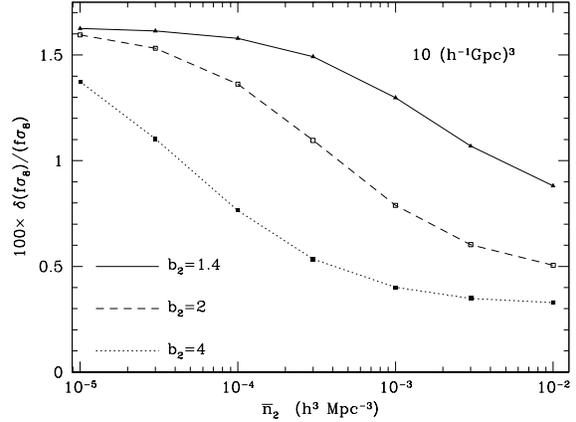}}
\end{center}
\caption{The fractional error on $f(z)\sigma_8(z)$ arising from a
$10\,(h^{-1}{\rm Gpc})^3$ survey at $z=0$ populated with two types of
galaxies.  The first population is held fixed with $b_1=1$ and
$\bar{n}_1=10^{-2}\,h^3\,{\rm Mpc}^{-2}$, i.e.~$\bar{n}P\gg 1$.
The second population has $b_2=1.4$ (solid), $b_2=2$ (dashed) or
$b_2=4$ (dotted) and the constraint is plotted vs.~$\bar{n}_2$.
All else being equal, the fractional constraints would be tighter at
higher $z$ where $f$ is larger.}
\label{fig:df_vs_nbar2}
\end{figure}

\subsubsection{The Fisher matrix}

The Fisher matrix for this problem is a simple generalization of
Eq.~(\ref{eqn:fisher}) but now there are $N(N+1)/2$ power spectra for
$N$ galaxy populations. For example, in the simplest case of two populations
there are 3 measured power spectra, which on large scales are estimates of
\begin{equation}
  P_{ij} = (b_i+f\mu^2)(b_j+f\mu^2)P_{\rm lin}
\end{equation}
where $i$ and $j$ run over $1$ and $2$ and $P_{12}=P_{21}$.

To calculate the Fisher matrix for multiple samples we need to sum
over $[N(N+1)/2]^2$ elements of the inverse covariance matrix
\begin{equation}
  F_{ij} = \sum_{XY}\int \frac{V_0\,d^3k}{(2\pi)^3}
  \ \left( \frac{\partial P_X}{\partial p_i} \right)
  C^{-1}_{XY}
  \left( \frac{\partial P_Y}{\partial p_j} \right)
  \quad ,
  \label{eq:fish_multiple}
\end{equation}
where we denote a pair of galaxy indices by $X$ or $Y$. Note that this
reduces to Eq.~(\ref{eqn:fisher}) for a single population. In order to
calculate the Fisher matrix, we need to determine the covariance
matrix and the derivatives of the power spectra with respect to the
cosmological parameters of choice.

\subsubsection{Calculating the covariance matrix}

If we assume that the bias is deterministic and that the shot-noise can be
treated as an (uncorrelated) Gaussian noise the covariance matrix for the
power spectra is straightforward to compute.
Similar results for the covariance matrix of quadratic combinations of
Gaussian fields have been determined previously when considering CMB
temperature and polarization power spectra
\citep[e.g.][]{Zaldarriaga97,Kamionkowski97} or the problem of combining
density and velocity power spectra \citep[e.g.][]{Burkey04}.
Our problem is slightly more general, in that we need to consider additional
combinations of power spectra, but similar in spirit.

If we define $N_a\equiv[1+1/(\bar{n}P_{aa})]$, the diagonal terms in
the covariance matrix are
\begin{eqnarray}
  \langle C_{aaaa} \rangle &=& 2P_{aa}^2N_a^2,\\
  \langle C_{abab} \rangle &=& P_{ab}^2 + P_{aa}P_{bb}N_aN_b,
\end{eqnarray}
where $a\ne b$, and the off-diagonal terms are calculated from
\begin{eqnarray}
  \langle C_{abcd} \rangle &=& 2P_{ab}P_{cd},
    \label{eq:cov_offdiag1} \\
  \langle C_{aabc} \rangle &=& 2P_{ab}P_{ac},
    \label{eq:cov_offdiag2} \\
  \langle C_{abac} \rangle &=& P_{ab}P_{ac} + P_{aa}P_{bc}N_a, 
    \label{eq:cov_offdiag3} \\
  \langle C_{aaab} \rangle &=& 2P_{ab}P_{aa}N_a,
    \label{eq:cov_offdiag4}
\end{eqnarray}
where $a\ne b\ne c\ne d$. These formulae are complete with the relations
$P_{ab}=P_{ba}$ and $\langle C_{XY}\rangle=\langle C_{YX}\rangle$.
The off-diagonal terms given in Eq.~(\ref{eq:cov_offdiag3}) do not occur in
the CMB example as there is only a single non-zero cross power there.  However,
all covariance matrix elements can be calculated using the same standard
procedure (see Appendix \ref{app:cov}).

\subsubsection{Calculating the derivatives}

If the parameters that we wish to constrain are $b_a$ and $f$ then
\begin{eqnarray}
  \frac{\partial P_{ab}}{\partial b_c} 
    &=& \left[(b_b+f\mu^2)\delta_K^{ac} + (b_a+f\mu^2)\delta_K^{bc}\right]
      P_{\rm lin},\\
  \frac{\partial P_{ab}}{\partial f} 
    &=& \left[(b_b+f\mu^2) + (b_a+f\mu^2)\right]
      \mu^2\ P_{\rm lin}
\end{eqnarray}
where $\delta_K^{ac}$ and $\delta^{bc}$ are Kronecker $\delta$s.  This completes
the input that we need for the Fisher matrix, Eq.~(\ref{eq:fish_multiple}).

We can include a parametrized line-of-sight smearing by multiplying the
power spectra by e.g.~$\exp[-(1/2)k^2\mu^2(\sigma_a^2+\sigma_b^2)]$.  The
derivatives are multiplied by the same factor and there is an additional
set
\begin{equation}
  \frac{\partial P_{ab}}{\partial \sigma^2_c} = -\frac{1}{2}k^2\mu^2
  \left( \delta_K^{ac}+\delta_K^{bc} \right) P_{ab}
\end{equation}

For the mode-by-mode parametrization developed in \S\ref{sec:mode} we can
use the logarithmic derivatives in Eq.~(\ref{eqn:mode_deriv}), multiplied
by $P_{\rm obs}(k_a,z_b)$.

\subsubsection{Results}

We confirm the finding of \citet{McDSel08} that using multiple populations
can result in significant gains in constraining power.  For example
Fig.~\ref{fig:df_vs_nbar2} shows the fractional error on $f$ decreases by a
factor of $2-3$ if a second population is simultaneously used to provide
constraints.  At fixed number density the gain is higher the more biased is
the second sample, and the constraint is weakened as the bias of the first
sample is increased.  Thus we would like to find a two samples with very
different clustering properties but reasonable number densities.

As long as the line-of-sight dispersion, $\sigma_i$, is not large the
marginalization has little effect on the total error.  One does, however,
prefer slightly higher $\bar{n}P$ when marginalizing over $\sigma_i$ than
when keeping it fixed.

The gains saturate quickly when using more than two samples.  In fact if
the total number of objects observed is to be held fixed, it is better to
increase the number densities of the lowest and highest biased sets rather
than include an intermediately biased sample at the expense of lower number
densities for all samples.

Within the deterministic bias model, splitting into multiple populations does
not affect constraints on the overall large-scale power spectrum shape: here
we are always limited by the total number of modes in the sample.
A bias model can be used to weight galaxies of different bias, allowing for
their different clustering strengths, in order to optimally calculate the
overall power spectrum shape \citep{PVP}. Any cosmological benefit from
splitting into multiple samples will therefore arise through better
constraints on $f(z)\sigma_8(z)$.

\section{Predictions for future surveys}

\begin{table}
\begin{center}
\begin{tabular}{lccc}
Survey & $\bar{n}$ & $z$  & $N_{\rm gal} $ \\
\hline
BOSS        & $3$ & $0.1<z<0.7$ & $1.5$ \\
WFMOS (1)   & $5$ & $0.5<z<1.3$ & $2.0$ \\
WFMOS (2)   & $5$ & $2.3<z<3.3$ & $0.6$ \\
EUCLID/JDEM &$50$ & $0.1<z<2.0$ & $500$
\end{tabular}
\end{center}
\caption{Fiducial parameters adopted as indicative of various planned or 
ongoing surveys.  $N_{\rm gal}$ is given in units of $10^6$, e.g.~BOSS has
1.5 million galaxies, and $\bar{n}$ in units of $10^{-4}\,h^3\,{\rm Mpc}^{-3}$.
For the survey with the proposed WFMOS instrument, we assume that this is
split into low (1) and high (2) redshift components as proposed in
\protect\citet{glazebrook05}.  We assume that each survey covers a fixed
fraction of the sky, so the volume within any redshift interval is completely
determined by these parameters.}
\label{tab:surveys}
\end{table}

In this section we apply our Fisher matrix formalism to 3 concepts for future
spectroscopic surveys\footnote{The HETDEX experiment \protect\citep{HETDEX}
has constraints similar to WFMOS(2) and we have not plotted it to avoid
crowding.}, with fiducial parameters given in Table~\ref{tab:surveys}.
We assume a tight prior on small-scale velocity dispersion.
The galaxy bias is one of the hardest parameters to predict for future surveys,
so we have adopted a conservative approach here.  We assume that redshift zero
galaxy bias is sampled from a uniform distribution with $1<b<2$.
The bias evolves with redshift such that the galaxy clustering amplitude is
constant.  For all surveys, we assume that we can use all modes with
$k<0.075\hompc$ at $z=0$, and that this limit evolves with redshift according
to the \citet{smith03} prescription for $k_{\rm nl}$.  This assumption is
deserving of further investigation in N-body simulations.

\begin{figure}
\begin{center}
\resizebox{3.1in}{!}{\includegraphics{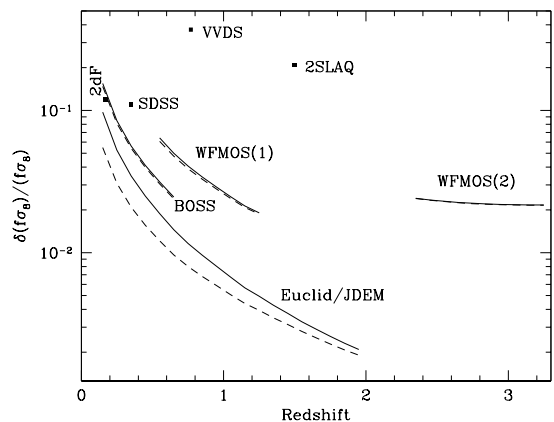}}
\end{center}
\caption{The fractional error on $f(z)\sigma_8(z)$ in bins of width
$\Delta z=0.1$, arising from fiducial surveys with parameters given in
Table~\protect\ref{tab:surveys}.  We consider all galaxies in a single bin
(solid lines), and split into 4 bins as a function of bias (dashed lines).
The existing constraints, as collected in \protect\citet{Song08b} and with
the addition of \protect\citet{Ang08}, are shown as solid squares (see text).}
\label{fig:df_vs_survey}
\end{figure}

Predictions for the error on $f(z)\sigma_8(z)$ are shown in
Fig.~\ref{fig:df_vs_survey}.
Results are presented either assuming that all galaxies are analyzed in a
single bin, or are split according to galaxy bias.
For surveys with a large number density, such as that proposed for the
EUCLID/JDEM concept, splitting into bins with different galaxy bias can
significantly reduce the expected errors, as we saw in
\S\ref{sec:fish_multiple}.
The galaxy sampling density proposed for the BOSS and WFMOS surveys is lower,
and we gain less from the multiple-sample approach.

It is useful to ask how these forecasts depend on the input assumptions.  As
an illustration, if we decrease the number density by a factor of $1.5$ the
BOSS, WFMOS(1) and unsplit Euclid/JDEM results are largely unchanged.
The error on WFMOS(2) increases by 30\% while the split Euclid/JDEM limit
increases by 10\%.
Decreasing the maximum $k$ from $0.075\hompc$ to $0.05\hompc$ at $z=0$ (with
the same scaling to higher $z$) all of the constraints become weaker.
The limit from BOSS increases by $\sim 40\%$, WFMOS(1) by $\sim 75\%$,
WFMOS(2) by $\sim 15\%$ and Euclid/JDEM by $\sim 85\%$ for the unsplit case
and $60\%$ for the split case.

We also include current constraints, as collected by \citet{Song08b} and
with the addtion of 2SLAQ, in Fig.~\ref{fig:df_vs_survey}.  These are
$\delta(f\sigma_8)/(f\sigma_8)=0.12$ at $z=0.12$ from the 2dFGRS
\citep{percival04}, $\delta(f\sigma_8)/(f\sigma_8)=0.11$ at $z=0.35$
from the SDSS LRG catalogue \citep{tegmark06},
$\delta(f\sigma_8)/(f\sigma_8)=0.37$ at $z=0.77$ from the VVDS \citep{guzzo08}
and $\delta(f\sigma_8)/(f\sigma_8)=0.21$ at $z=1.5$ from 2SLAQ \citep{Ang08}.
As can be seen, the next generation of spectroscopic galaxy surveys will provide
an order of magnitude increase in the available cosmological constraints from
redshift-space distortions at $z>0.1$.

\begin{figure}
\begin{center}
\resizebox{3.1in}{!}{\includegraphics{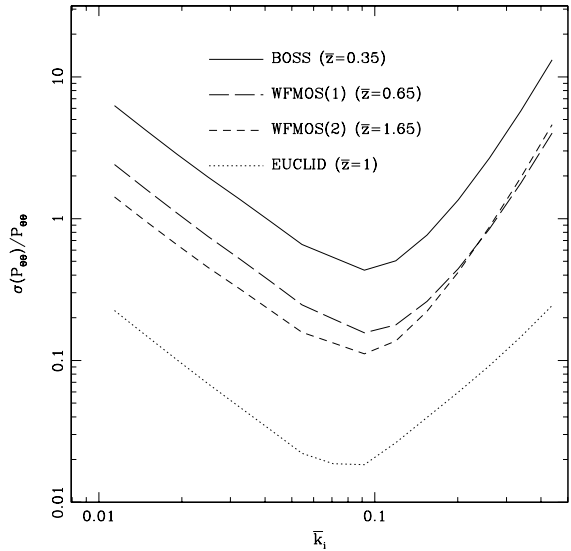}}
\end{center}
\caption{The fractional error on $P_{\Theta\Theta}$ using the experimental
parameters in Table \protect\ref{tab:surveys}.  We show only a single,
representative bin in redshift for each experiment, to avoid clutter.}
\label{fig:dPk_ex}
\end{figure}

The direct constraint on $P_{\Theta\Theta}$ for our futuristic surveys is
shown in Fig.~\ref{fig:dPk_ex}, for some representative bins in redshift.
We expect to be able to place tight constraints on the velocity power
spectrum near $k\simeq 0.1\,h\,{\rm Mpc}^{-1}$ with future experiments.

\section{Conclusions} \label{sec:conclusions}

Observations of redshift-space distortions in spectroscopic galaxy surveys
offer a powerful way to measure the large-scale velocity field, which in
turn provides a sensitive test of both the expansion rate of the Universe
and our theory of gravity.  We have developed a Fisher matrix formalism
which allows forecasting of the constraints future, idealized, surveys
would be able to place on the linear growth rate,
$f(z)\sigma_8(z)\propto dD/d\ln a$,
and shown that they are potentially highly constraining, though not as
constraining as the scaling of \citet{guzzo08} predicts.

We have developed the Fisher matrix exposition in multiple levels of
sophistication and realism, assuming strict functional forms for the
power spectra or allowing them to float freely.  As expected the constraints
are tightest when theoretical investigations can provide good priors for the
form and range of parameters, but even relatively conservative assumptions
suggest that percent level measurements of $f$ should be possible with future
surveys.
Further work on understanding the correlation of velocity and density fields,
scale-dependent bias and non-linear effects could pay big dividends.

As pointed out by \citet{McDSel08}, using multiple populations of galaxies
can tighten the constraint on $f(z)\sigma_8(z)$ (though it does not improve
measurement of the shape of $P_{\Theta\Theta}$).
We show that this can be naturally incorporated into our formalism.
The largest improvement comes when disjoint samples with a large difference
in bias, both having a high number density
[$\bar{n}P(k\simeq 0.1\,h\,{\rm Mpc}^{-1})\gg 1$] are used.
The ultimate limit to this method will come from stochasticity in the biasing
of galaxies, and which types of galaxies minimize this effect on which scales
is an important avenue for further investigation.

We have made a number of simplifications in this analysis which it will be
important to address in future work.  First we have assumed that the
large-scale velocity field of the galaxies is that of the matter.  Ultimately
our ability to model any velocity bias will set a lower limit on what can
be achieved.  It is important to note that we are limited by how accurately
the halo centers follow the mass velocity field ``on large scales'', rather
than a bias within the halos.  For the former, simulations suggest that halos
do tend to trace the mass very well \citep{HSWSW07,PerWhi08}.
Current observational constraints on the latter from modeling clusters are
consistent with no velocity bias at the $10\%$ level \citep[e.g.][]{SCSM}.
Simulations suggest little or no velocity bias for the majority of
``galaxies'' in dark matter \citep{Spr01,FalDie06} and hydrodynamic
\citep{Ber03} simulations at the same level.
Investigations of these phenomena in simulations of the standard cosmology
and with alternative theories of gravity will be very valuable.

A code to compute the Fisher matrix given survey parameters is available
at {\tt http://mwhite.berkeley.edu/Redshift}.

\section*{Acknowledgments}

MW thanks Uros Seljak and YS thanks Olivier Dore for discussions on redshift
space distortions.
MW is supported by NASA.
YS is supported by STFC.
WJP is supported by STFC, the Leverhulme Trust and the European Research
Council.
The simulations used in this paper were analyzed at the National Energy
Research Scientific Computing Center.

\appendix

\section{The off-diagonal power spectrum components} \label{app:cov}
  
In this section, we derive formulae for $\langle C_{aabc}\rangle$, and
$\langle C_{abac}\rangle$, as given in Eqns.~\ref{eq:cov_offdiag2}
\&~\ref{eq:cov_offdiag3}. Similar derivations for CMB power spectra were
presented in \citet{Kamionkowski97}.

Suppose that, in a particular experiment, we have $M$ independent
complex samples ${\bf \delta}^m$, with $1<m<M$, drawn from multivariate
Gaussian distribution with $\hat{P}_{ab}=1/M\sum_m\langle
\delta^{m*}_a\delta_b^{m}\rangle$. Our estimate of the covariance
between power spectrum measurements from this experiment is
\begin{equation}
  \langle\hat{P}_{ab}\hat{P}_{cd}\rangle = \frac{1}{M}\sum_{m,m'}
    \langle \delta^{m*}_a\delta^{m}_b\delta^{m'*}_c\delta^{m'}_d\rangle 
    \label{eq:cov_full}
\end{equation}
To proceed, we split this sum into terms with $m=m'$ and $m\ne
m'$. Where $m=m'$, we can use the standard result for the 4-order
moments of multivariate Gaussian random variables that, if $x_a$ are
real and Gaussian distributed, the expectation
\begin{equation}
  E[x_ax_bx_cx_d]=\langle x_ax_b\rangle \langle x_cx_d\rangle
    +\langle x_ax_d\rangle \langle x_bx_c\rangle
    +\langle x_ax_c\rangle \langle x_bx_d\rangle.
\end{equation}
For the component of Eq.~\ref{eq:cov_full} where $m\ne m'$, we can
easily decompose into 2-order moments. For
$\langle\hat{P}_{aa}\hat{P}_{bc}\rangle$ this procedure gives
\begin{eqnarray}
  \langle\hat{P}_{aa}\hat{P}_{bc}\rangle 
    &=& \frac{1}{M^2}\sum_{m,m'}
    \left[\langle|\delta^m_a|^2\delta^{m*}_b\delta^{m}_c\rangle\delta_K^{mm'}+
    \right. \nonumber \\
    && \left.
      \langle|\delta^m_a|^2\delta^{m'*}_b\delta^{m'}_c\rangle(1-\delta_K^{mm'})\right]\\
    &=& \frac{1}{M}\left[\hat{P}_{aa}\hat{P}_{bc} +2\hat{P}_{ab}\hat{P}_{ac}\right]
      +\hat{P}_{aa}\hat{P}_{bc}-\frac{1}{M}\hat{P}_{aa}\hat{P}_{bc}\\
    &=& \frac{2}{M}\hat{P}_{ab}\hat{P}_{ac}+\hat{P}_{aa}\hat{P}_{bc},
\end{eqnarray}
so $\langle C_{aabc}\rangle=2\hat{P}_{ab}\hat{P}_{ac}$ divided by the number of modes.

To calculate $C_{abac}$, note that this procedure also gives that
$\langle\hat{P}_{ab}\hat{P}_{ac}\rangle =
1/M[\hat{P}_{ab}\hat{P}_{ac}+\hat{P}_{aa}\hat{P}_{bc}] + \hat{P}_{ab}\hat{P}_{ac}$, 
so $\langle C_{abac}\rangle=\hat{P}_{aa}\hat{P}_{bc}+\hat{P}_{ab}\hat{P}_{ac}$, 
divided by the number of modes.
The other terms in the covariance matrix can be calculated using
the same methodology.

\label{lastpage}
\end{document}